\documentclass[sn-mathphys,Numbered,12pt]{sn-jnl}
\usepackage{graphicx}
\usepackage{multirow}
\usepackage{amsmath,amssymb,amsfonts}
\usepackage{amsthm}
\usepackage{mathrsfs}
\usepackage[title]{appendix}
\usepackage{xcolor}
\usepackage{textcomp}
\usepackage{manyfoot}
\usepackage{booktabs}
\usepackage{algorithm}
\usepackage{algorithmicx}
\usepackage{algpseudocode}
\usepackage{listings}

\theoremstyle{thmstyleone}
\newtheorem{theorem}{Theorem}
\newtheorem{proposition}[theorem]{Proposition}
\theoremstyle{thmstyletwo}
\newtheorem{lemma}{Lemma}
\newtheorem{corollary}{Corollary}
\newtheorem{example}{Example}
\newtheorem{remark}{Remark}

\newtheorem{note}{Note}

\theoremstyle{thmstylethree}
\newtheorem{definition}{Definition}

\begin{document}
	
	\title[Article Title]{On equidistant single-orbit cyclic and quasi-cyclic subspace codes}

	\author{\fnm{Mahak} }\email{mahak@ma.iitr.ac.in}
	
	\author*{\fnm{Maheshanand} \sur{Bhaintwal}}\email{maheshanand@ma.iitr.ac.in}

	\affil{\orgdiv{Department of Mathematics}, \orgname{Indian Institute of Technology Roorkee}, \orgaddress{\street{} \city{Roorkee}, \postcode{247667}, \state{Uttarakhand}, \country{India}}}

	\abstract{A code is said to be equidistant if the distance between any two distinct codewords of the code is the same. In this paper, we have studied equidistant single-orbit cyclic and quasi-cyclic subspace codes. The orbit code generated by a subspace $U$ in $\mathbb{F}_{q^n}$ such that the dimension of $U$ over $\mathbb{F}_q$ is $t$ or $n-t$, $\mbox{where}~t=\dim_{\mathbb{F}_q}(\mbox{Stab}(U)\cup\{0\})$, is equidistant and is termed a trivial equidistant orbit code. Using the concept of cyclic difference sets, we have proved that only the trivial equidistant single-orbit cyclic subspace codes exist. Further, we have explored equidistant single-orbit quasi-cyclic subspace codes, focusing specifically on those which are sunflowers.}


	\maketitle
	
	\section{Introduction}\label{sec1}
	
	Subspace codes are used in random network coding to correct errors and erasures. A well-known paper \cite{koetter} by K\"{o}tter and Kschischang sparked the main interest in subspace codes. Since $2008$, researchers have been actively engaged in working on these codes. Of special interest among subspace codes is the class of cyclic subspace codes, introduced by Etzion and Vardy \cite{vardy}. The algebraic structure of cyclic subspace codes and their efficient encoding and decoding algorithms motivate the study of these codes. 
	
	A cyclic subspace code $C$ is a  collection of $\mathbb{F}_q$-subspaces of $\mathbb{F}_{q^n}$ that is closed under the multiplication by the elements of $\mathbb{F}_{q^n}^*$, i.e., $\alpha U\in C$ for all $U\in C$ and $\alpha\in \mathbb{F}_{q^n}^*$. In \cite{traut}, it is shown that the set $\{\alpha U\mid \alpha \in \mathbb{F}_{q^n}^*\}$ can be seen as the orbit of a subspace $U$ under the action of the group $\mathbb{F}_{q^n}^*$ on the set of subspaces of $\mathbb{F}_{q^n}$. Thus, a cyclic subspace code is the union of the orbits of the subspaces contained in it. The code $\mbox{Orb}(U)=\{\alpha U \mid \alpha \in \mathbb{F}_{q^n}^*\}$ is called a single-orbit cyclic subspace code. Quasi-cyclic subspace codes are a natural generalization of cyclic subspace codes and are studied in \cite{glue}. A subspace code $C$ is called a quasi-cyclic subspace code if $\alpha U\in C $ for all $U\in C $ and $\alpha \in G$, where $G$ is a multiplicative subgroup of $\mathbb{F}_{q^n}^*$ (see \cite{otal}). 
	
	   A code with the property that the distance between any two distinct codewords is the same is called an equidistant code. Equidistant subspace codes have been explored by many researchers (see \cite{tuvi,bartoli,elisa,basu}).  The applications of equidistant subspace codes to distributed storage systems are discussed in \cite{raviv}.
	   
In the literature, a significant amount of research has been done on  single-orbit cyclic subspace codes. For a subspace $U$ of dimension $k~(<n)$ in $\mathbb{F}_{q^n}$, the subspace distance of $\mbox{Orb}(U)$ is $2k-2l$ for some $l,~0\leq l\leq k-1$. The first case of orbit codes that was studied was of spread codes \cite{MGR2008}, which corresponds to $l=0$, i.e., the subspace distance of $\mbox{Orb}(U)$ is $2k$. If $k$ divides $n$, we can construct a single-orbit cyclic subspace code with minimum distance $2k$ and cardinality $\frac{q^n-1}{q^k-1}$ for any $q$. In  \cite{traut}, the authors studied the case $l=1$, i.e., the subspace distance of $\mbox{Orb}(U)$ is $2k-2$.  They focused on the codes $\mbox{Orb}(U)$ of maximum cardianlity $\frac{q^n-1}{q-1}$ with the subspace distance $2k-2$, and referred to these codes as optimal full-length single-orbit codes. They conjectured the existence of such codes for any possible values of $q, n, \mbox{and}~ k$. Some years later, in \cite{RRT2017} Roth, Raviv and Tamo solved the conjecture for most of the cases by observing that constructing a single-orbit cyclic subspace code of maximum size $\frac{q^n-1}{q-1}$ and minimum distance $2k-2$ is equivalent to constructing a Sidon space of dimension $k$ in $\mathbb{F}_{q^n}$. A subspace $U$ in $\mathbb{F}_{q^n}$ is called a \emph{Sidon} space if for all non-zero $a,b,c,d \in U $,  $ab=cd$ implies that $\{a \mathbb{F}_q, b\mathbb{F}_q\}=\{c\mathbb{F}_q, d\mathbb{F}_q\}$, where $e\mathbb{F}_{q}$ stands for the set $\{\lambda e: \lambda \in \mathbb{F}_q\}$ for any $e\in \mathbb{F}_{q^n}$.
	Sidon spaces were introduced in \cite{BSZ2017} and have been extensively studied  over the years. In particular, many constructions of Sidon spaces have been proposed, including those for constructing optimal multi-orbit codes; see \cite{FW2021,ZC2022,ZG2022,ZT2023,ZT2023-2,ZTH2023,LL2023,CPSZ2023,Z2023,ZTC2024,YJ2024, HC2024,C2025}.

In \cite{leh}, the authors investigated quasi-optimal codes, i.e., the case of minimum distance $2k-4$. In the same paper, Gluesing-Luerssen and Lehmann introduced the notions of the weight spectrum and weight distribution of a single-orbit cyclic subspace code. In this paper, we study the single-orbit cyclic subspace codes for which the minimum distance coincides with the distance between any two distinct codewords. Thus, the weight spectrum of such codes consists of only one value. 
	The opposite perspective was explored in \cite{CPZ2024-2}, where the authors analyzed the case in which the weight spectrum of $\mbox{Orb}(U)$ is complete, i.e., for any $i=1,\ldots,k$, there exist $V,W\in \mbox{Orb}(U)$ such that the subspace distance between $V$ and $W$ is $2i$. 
		In \cite{CPZ2024}, the authors have completely classified the single-orbit cyclic subspace codes of dimension $3$. By this classification (see \cite[Corollary III.8]{CPZ2024}), it immediately follows that there do not exist any
	non-trivial equidistant single-orbit codes, $\mbox{Orb}(U)$, when $k=3$. In this paper, we prove that this is true for any $k\geq 2$, using a different approach based on cyclic difference sets.

	This paper focuses on equidistant single-orbit cyclic and quasi-cyclic subspace codes. A code $\mbox{Orb}(U)$ is always equidistant if the dimension of $U$ is $t$ or $n-t$ over $\mathbb{F}_q$, where $t=\dim_{\mathbb{F}_q}(\mbox{Stab}(U)\cup\{0\})$. We call such codes trivial equidistant orbit codes. In Section 2, we introduce the fundamentals of subspace codes and cyclic difference sets. In Section 3, with the help of cyclic difference sets, we prove  that there are only trivial equidistant single-orbit cyclic subspace codes. In Section 4, we provide some examples to illustrate the existence of single-orbit quasi-cyclic subspace codes. We introduce sunflower orbit codes and prove that $\mbox{Orb}_{\beta}(U)$ is always a  sunflower code if $\beta$ is an element of degree $2$ in $\mathbb{F}_{q^n}$. Moreover, we discuss the maximum cardinality of sunflower codes.

	\section{Preliminaries}\label{sec2}

Let $q=p^h,~ p$ a prime and $h$ a positive integer. Let $\mathbb{F}_q$ be a finite field of size $q$, and let $\mathbb{F}_q^*:= \mathbb{F}_q\backslash \{0\}$. Let $\mathbb{F}_{q^n}$ denote the extension field of degree $n$ of $\mathbb{F}_q$, and let $\mathbb{F}_{q^n}^*:= \mathbb{F}_{q^n} \backslash \{0\}$. For $\beta\in \mathbb{F}_{q^n}^*$, let $\lvert\beta\rvert$ denote the order of $\beta$ in the multiplicative group $(\mathbb{F}_{q^n}^*,\times)$. For any subset $\{x_1,x_2,\ldots,x_r\}$ of $\mathbb{F}_q^n$ the
subspace of $\mathbb{F}_q^n$ over $\mathbb{F}_q$ spanned by this set is denoted by $\langle x_1,x_2,\ldots,x_r\rangle_{\mathbb{F}_q}$. For any element $\alpha\in \mathbb{F}_{q^n}$, we define $\overline{\alpha} :=\alpha\mathbb{F}_q=\{\alpha\lambda\mid \lambda\in \mathbb{F}_q\}$. 

The set of all $\mathbb{F}_q$-subspaces of $\mathbb{F}_{q}^n$ is denoted by $\mathcal{P}_q(n)$ and is called the projective space of order $n$ over $\mathbb{F}_q$. For $0\leq k\leq n$, the Grassmanian of dimension $k$, denoted by $\mathcal{G}_q(n,k)$, is the set of all $k$-dimensional $\mathbb{F}_q$-subspaces of $\mathbb{F}_q^n$. Clearly, 
\[\mathcal{P}_q(n)= \bigcup_{k=0}^{n}\mathcal{G}_q(n,k)~.\]
The size of $\mathcal{G}_q(n,k)$ is given by the $q$-binomial coefficient ${n \brack k}_q$, i.e., 
\[ |\mathcal{G}_q(n,k)|=  {n \brack k}_q=  \frac{(q^n-1) (q^n-q)\ldots(q^n-q^{k-1})} {(q^k-1) (q^k-q)\ldots(q^k-q^{k-1})}~.\]

The projective space $\mathcal{P}_q(n)$ is a metric space with respect to the metric  $d_s$, defined by,
\[d_s(U,V)=\dim(U)+\dim(V)-2 \dim(U\cap V)\]
for any $U,V\in \mathcal{P}_q(n)$. 
A subspace code $C$ is a subset of $\mathcal{P}_q(n)$ containing at least two elements, with the metric $d_s$.  The minimum distance of a subspace code $C$, denoted by $d_s(C)$, is defined by
\[d_s(C) = \mbox{min}\{d_s(U,V)\mid U,V \in C,~ U\neq V\}~.\]
The subspace code $C$ is said to be a constant dimension subspace code if every element in  $C$ is of the same dimension, i.e., $C\subseteq \mathcal{G}_q(n,k)$ for some positive integer $k\leq n$.
For a constant dimension subspace code $C\subseteq \mathcal{G}_q(n,k)$, we have 
\[d_s(C)=2k-2~\mbox{max}\{\dim(U\cap V) \mid U,V\in C,~ U\neq V\}~.\]
A  constant dimension subspace code $C$ is said to be an\emph{ equidistant subspace code} if for all $U,V\in C$ with $U\neq V$ we have $d_s(U,V)=d_s(C)$. An equidistant subspace code $C\subseteq \mathcal{G}_q(n,k)$ is said to be $c$-intersecting if $d_s(C)=2(k-c)$, i.e., $\dim(U\cap V)=c$ for all $U,V\in C$ with $U\neq V$. 

It is well known that the extension field $\mathbb{F}_{q^n}$ is a vector space of dimension $n$ over $\mathbb{F}_q$, and both $\mathbb{F}_q^n$ and $\mathbb{F}_{q^n}$ are isomorphic as a vector space over $\mathbb{F}_q$. Due to the rich algebraic structure of $\mathbb{F}_{q^n}$,  compared to that of $\mathbb{F}_q^n$,  we identify the subspaces of $\mathbb{F}_q^n$ with those of  $\mathbb{F}_{q^n}$ in the study of cyclic subspace codes. 

For a subspace $U\subseteq \mathbb{F}_{q^n} ~\mbox{and} ~\alpha \in \mathbb{F}_{q^n}^*$, the cyclic  shift of $U$ with respect to $\alpha$ is defined as $\alpha U=\{\alpha u \mid u \in U \}$. Clearly $\alpha U \subseteq \mathbb{F}_{q^n}$. It is easy to see that $\alpha U$ is a vector space over $\mathbb{F}_q$ and its dimension is the same as the dimension of $U$ over $\mathbb{F}_q$. In fact, we can define a group action $\mathbb{F}_{q^n}^* \times \mathcal{P}_q(n) \rightarrow \mathcal{P}_q(n)$ of $\mathbb{F}_{q^n}^*$ on $\mathcal{P}_q(n)$ (see \cite{traut}) as 
\begin{eqnarray*}
	(\alpha , U) &\rightarrow & \alpha U~.
\end{eqnarray*}
For any $\mathbb{F}_q$-subspace $U \subseteq \mathbb{F}_{q^n}$, the orbit of $U$, denoted by $\mbox{Orb}(U)$, is defined by \[\mbox{Orb}(U)=\{\alpha U \mid \alpha \in  \mathbb{F}_{q^n}^* \}~.\]
The \emph{stabilizer} of $U$, denoted by $\mbox{Stab}(U)$, is defined by $\mbox{Stab}(U)=\{\alpha \in \mathbb{F}_{q^n}^*\mid \alpha U=U\}$. Clearly, $\mbox{Stab}(U)$ is a subgroup of $\mathbb{F}_{q^n}^*$, and since $aU=U$ for all $a \in \mathbb{F}_q^\ast$, we have $\mathbb{F}_q^\ast \subseteq \mbox{Stab}(U)$. By  \cite[Lemma 3.3]{glue}, $\mbox{Stab}(U) \cup \{0\}$ is a subfield of $\mathbb{F}_{q^n}$, and $U$ is a vector space over $\mbox{Stab}(U)\cup\{0\}$. Thus, $\mbox{Stab}(U)\cup \{0\}= \mathbb{F}_{q^t}$ for some $t$, which is clearly a divisor of $\gcd(\dim_{\mathbb{F}_q}(U),n)$. For $t=n$, i.e., $U=\mathbb{F}_{q^n}$, we have $\mbox{Stab}(U)=\mathbb{F}_{q^n}^*$. Thus, in this case, $\mbox{Orb}(U)$ contains only one element. So, we always consider $t<n$. By \cite[Theorem 1]{otal}, for any subspace $U$ of $\mathbb{F}_{q^n}$, we have
\[\lvert \mbox{Orb}(U)\rvert = \frac{q^n-1}{\lvert \mbox{Stab}(U)\rvert} = \frac{q^n-1}{q^t-1}~.\]
A subspace code $C$ with the property that for any $\alpha \in \mathbb{F}_{q^n}^*~\mbox{and}~U\in C,~\alpha U\in C$, is said to be a \emph{cyclic subspace code}. Thus, $\mbox{Orb}(U)$ is a cyclic constant dimension subspace code, and we call $\mbox{Orb}(U)$ a \emph{single-orbit cyclic subspace code} or simply an \emph{orbit code}. In general, a cyclic subspace code is the union of the orbits under the action of  $\mathbb{F}_{q^n}^*$ of the subspaces.

If $\mbox{Stab}(U)=\mathbb{F}_q^*$, i.e., $\lvert{\mbox{Orb}(U)}\rvert= \frac{q^n-1}{q-1}$, then $\mbox{Orb}(U)$ is called a \emph{full-length orbit code} and we say that $U$ generates a full-length orbit. Otherwise, $\mbox{Orb}(U)$ is a degenerate orbit.

Let $U$ be an $\mathbb{F}_q$-subspace of dimension $k$ in $\mathbb{F}_{q^n}$.  By the definition of subspace distance, for any $\alpha, \beta \in \mathbb{F}_{q^n}^*$,  we have 
\begin{eqnarray*}
	d_s(\alpha U, \beta U)&=&\dim(\alpha U)+\dim(\beta U)-2 \dim(\alpha U \cap \beta U)\\
	&=& 2k - 2 \dim(\alpha U \cap \beta U)~.
\end{eqnarray*}
As, $\dim(\alpha U\cap\beta U)= \dim(U\cap\alpha^{-1} \beta U)$, we get
\[d_s(\alpha U, \beta U)= 2k- 2\dim(U\cap \alpha^{-1}\beta U) ~.\]
Therefore,
\begin{eqnarray*}
	d_s(\mbox{Orb}(U))&=& \mbox{min}\{d_s(\alpha U,\beta U)\mid \alpha ,\beta \in \mathbb{F}_{q^n}^*, \alpha U \neq \beta U \}\\
	&=& 2k- 2~\mbox{max}\{\dim(U\cap\gamma U)\mid \gamma \in \mathbb{F}_{q^n}^*, \gamma U \neq U\}~.
\end{eqnarray*}
Thus, for any subspace $U$ in $\mathbb{F}_{q^n}$, $\mbox{Orb}(U)$ is $c$-intersecting equidistant subspace code if there exists some non-negative integer $c$ such that $\dim(U\cap \alpha U)=c$  for all $\alpha\in \mathbb{F}_{q^n}^*\backslash \mbox{Stab}(U)$. If $\mbox{Orb}(U)$ is an equidistant code, then we call it \emph{equidistant single-orbit cyclic subspace code}.

Next, we discuss the cases in which $\mbox{Orb}(U)$ is trivially an equidistant code.

	\noindent\textbf{Case 1:} Suppose $\dim_{\mathbb{F}_q}(U)=t$ where $\mbox{Stab}(U) = \mathbb{F}_{q^t}^*$ and $t \geq 1,~ t\mid n$. Let $\alpha \in\mathbb{F}_{q^n}\backslash \mathbb{F}_{q^t}$. Note that $U,~ \alpha U$ and hence $U \cap \alpha U$ are $ \mathbb{F}_{q^t}$-subspaces. Thus, 
	\[0 \leq \dim_{\mathbb{F}_{q^t}}(U \cap \alpha U) \leq 1 = \dim_{\mathbb{F}_{q^t}}(U)~,\]
	and since $ \alpha \notin \mbox{Stab}(U),~ \dim_{\mathbb{F}_q}
	(U \cap \alpha U) = 0$ for any $\alpha \in \mathbb{F}_{q^n}\backslash \mathbb{F}_{q^t}$. Thus, $\mbox{Orb}(U)$ is a $0$-intersecting equidistant subspace code.\\
	\textbf{Case 2:} Suppose $\dim_{\mathbb{F}_q}(U)=n-t$ where $\mbox{Stab}(U) = \mathbb{F}_{q^t}^*$ and $t \geq 1,~ t\mid n$. Let $n =tl~ \mbox{for some}~ l \in \mathbb{Z}$. For any $\alpha \in\mathbb{F}_{q^n}\backslash \mathbb{F}_{q^t}$, by the Grassmann formula
	we get
	\[t(l-2) \leq  \dim_{\mathbb{F}_q}(U \cap \alpha U) \leq t(l-1)~.\]
	Since $U \cap \alpha U$ is an $\mathbb{F}_{q^t}$-subspace, we get $\dim_{\mathbb{F}_q}(U \cap \alpha U) = t(l-2) = n-2t~ \mbox{
		for any}~ \alpha \in \mathbb{F}_{q^n}\backslash \mathbb{F}_{q^t}$. So, in this case, $\mbox{Orb}(U)$ is a $(n-2t)$-intersecting equidistant subspace code.

Hence, if $\dim(U)=t$ or $n-t$ and $\mbox{Stab}(U)=\mathbb{F}_{q^t}^*$, where $t\geq 1$ and $t\mid n$, then $\mbox{Orb}(U)$ is an equidistant subspace code. We call this code a trivial equidistant single-orbit cyclic subspace code. Therefore, from now on, we will assume that $\dim(U)>1$ and  $\mbox{Stab}(U)=\mathbb{F}_{q^t}^*$ for some $t<\dim(U)$.

\begin{definition}\cite[Definition 3.1]{stin}
	Suppose $(G,+)$ is a finite group of order $v$ in which the identity element is denoted by  $``0"$. Let $k$ and $\lambda$ be positive integers such that $2\leq k<v$. A $(v,k,\lambda)$-difference set in $(G,+)$ is a subset $D\subseteq G$ that satisfies the following properties: 
	\begin{enumerate}
		\item $\lvert D\rvert =k$,
		\item the multiset $[x-y: x,y \in D, x\neq y]$ contains every element in $G\backslash \{0\}$ exactly $\lambda$ times.
	\end{enumerate}
\end{definition} 
Note that, if a $(v,k,\lambda)$-difference set exists, then
\begin{equation}\label{eqn1}
	\lambda(v-1)=k(k-1)~.
\end{equation}
Let $D$ be a $(v,k,\lambda)$-difference set in a group $(G,+)$. For any $g\in G$, define \[D+g=\{x+g: x \in D\}~.\] Any set $D+g$ is called a translate of $D$. 
\begin{lemma}\cite[p. 372]{van}\label{lemma1}
	Let $(G,+)$ be a group of order $v$ and $D\subseteq G$ be a $(v,k,\lambda)$-difference set. Then for any $g,g'\in G,~ g \neq g'$, we have $\lvert (D+g)\cap (D+g')\rvert =\lambda$. 
\end{lemma}

\begin{definition}\cite[Definition 2.3]{jun}
	Let $(G,+)$ be a group of order $nm$ and let $(N,+)$ be a subgroup of $G$ of order $n$. Then a $k$-subset $D$ of $G$ is called a relative difference set with parameters $n,m,k,\lambda_1$ and $\lambda_2$ (relative to $N$) or briefly an $(n,m,k,\lambda_1,\lambda_2)$-RDS, provided that  the list of differences $\{d_1-d_2: d_1,d_2\in D, d_1\neq d_2\}$ contain each element of $N$, except zero, precisely $\lambda_1$ times and each element of $G\backslash N$ exactly $\lambda_2$ times. 
\end{definition}

\begin{lemma}\cite[Lemma 2.5]{jun}\label{lemma2}
	Let $D$ be an $(n,m,k,\lambda_1,\lambda_2)$-RDS in $G$. Then 
	\[k(k-1)=n(m-1)\lambda_2+(n-1)\lambda_1~.\]
\end{lemma}
A difference set or relative difference set is said to be \emph{cyclic} if $G$ is a cyclic group.

In \cite{ghat}, the author established a connection between the construction of cyclic orbit codes with a given minimum distance and cyclic difference sets, and  stated the following proposition.
\begin{proposition}\cite[Proposition 4]{ghat}
	Let $\alpha \in \mathbb{F}_{q^n}^*$ be a primitive element over $\mathbb{F}_q$, and let $u_{\alpha}\in \mathbb{F}_q[X]$ be the minimal polynomial of $\alpha$ over $\mathbb{F}_q$. Consider the orbit of the $k$-dimensional subspace $U=\{0, \alpha^{i_1},\ldots,\alpha^{i_{q^k-1}}\}, i_j\in \mathbb{Z}_{q^n-1}$ for all $j=1,\ldots, q^k-1$ under the action of the Singer subgroup generated by $C_{u_\alpha}$, where $C_{u_\alpha}$ is the companion matrix of the polynomial $u_\alpha$. If the indices $i_j$ constitute a $(v=q^n-1,\ell=q^k-1,\lambda)$ difference set, where $\lambda\leq q^d-1$, then the orbit code so formed has minimum subspace distance $2(k-d)$.  
\end{proposition}

\section{Equidistant single-orbit cyclic subspace codes}\label{sec3}
We first discuss $0$-intersecting equidistant single-orbit cyclic subspace codes. We will show that non-trivial $0$-intersecting equidistant single-orbit cyclic subspace  code does not exist. For this, we define the spread and partial spread codes. 
\begin{definition}\cite[Definition 21]{seg}\cite[4.1]{hir}
	For any $k~(<n)$, a $k$-spread  is a collection of $k$-dimensional subspaces $\{X_1,X_2,\ldots,X_t\}$ of $\mathbb{F}_q^n$ such that
	\begin{enumerate}
		\item $X_i \cap X_j=\{0\}$, for $i\neq j, 1\leq i,j\leq t.$
		\item $\bigcup\limits_{i=1}^{t} X_i =\mathbb{F}_q^n$.
	\end{enumerate}
\end{definition}
\begin{theorem}\cite[Theorem 22]{seg}\cite[Theorem 4.1.1]{hir}\label{thm1}
	A $k$-spread exists if and only if $k$ divides $n$. Moreover, the cardinality of a $k$-spread is $\frac{q^n-1}{q^k-1}$.
\end{theorem}

\begin{definition}\cite[Definition 6]{gorla}
	A partial $k$-spread of $\mathbb{F}_q^n$ is a subset $\mathcal{A}\subseteq \mathcal{G}_q(n,k)$ such that $U\cap V=\{0\}$ for all $U, V\in \mathcal{A}$ with $U\neq V$. A partial $k$-spread of $\mathbb{F}_q^n$  with at least two elements is a $q$-ary subspace code of length $n$, dimension $k$ and minimum distance $2k$. We call such a code a partial spread code
\end{definition}
\begin{lemma}\cite[Lemma 7]{gorla}\label{lem5}
	Let $\mathcal{A}\subseteq\mathcal{G}_q(n,k)$ be a partial $k$-spread code. Denote by $r$ the remainder obtained when $n$ is divided by $k$. Then 
	\[ \lvert\mathcal{A}\rvert\leq \frac{q^n-q^r}{q^k-1}~.\]
\end{lemma}

\begin{remark}\label{newre}
Let $U$ be a subspace in $\mathbb{F}_{q^n}$ of dimension $k (\geq 2)$, and let $\alpha$ be a primitive element of $\mathbb{F}_{q^n}$.  Suppose that $\mbox{Orb}(U)$ is a $0$-intersecting equidistant code, i.e., $\dim(U\cap\alpha^i U)=0$  for all $i,~ 1\leq i\leq q^n-1$, whenever $\alpha^i U\neq U$. First, consider  $U$ generates a full-length orbit. Then, $\lvert\mbox{Orb}(U)\rvert =\frac{q^n-1}{q-1}$.\\ 
\textbf{Case 1}. Let $k$ divide $n$. By Theorem \ref{thm1},
$\mbox{Orb}(U)$ is a subset of $k$-spread and
\[\lvert\mbox{Orb}(U)\rvert\leq \frac{q^n-1}{q^k-1}~.\]
Since $k\geq 2$, this is not possible. So, in this case, $\mbox{Orb}(U)$ cannot be a $0$-intersecting equidistant code. \\
\textbf{Case 2}. Let $k$ does not divide $n$, and let $r$ denote the remainder obtained when $n$ is  divided by $k$. Then, $\mbox{Orb}(U)$ is a subset of partial $k$-spread. Therefore, by Lemma \ref{lem5}, we get 
\[\lvert\mbox{Orb}(U)\rvert \leq \frac{q^n-q^r}{q^k-1}~.\]

Clearly, $\lvert \mbox{Orb}(U)\rvert= \frac{q^n-1}{q-1} >\frac{q^n-q^r}{q^k-1}$. Thus, for this case also, $\mbox{Orb}(U)$ cannot be a $0$-intersecting equidistant code. \\
Now, suppose that $U$ does not generate a full-length orbit. Let $\mbox{Stab}(U)=\mathbb{F}_{q^t}^*$, where $2\leq t <k$ and $t$ is a divisor of $\mbox{gcd}(k,n)$. Thus, $\lvert\mbox{Orb}(U)\rvert=\frac{q^n-1}{q^t-1}$. Let $r$ denote the remainder obtained when $n$ is  divided by $k$. Observe that $\lvert\mbox{Orb}(U)\rvert =\frac{q^n-1}{q^t-1}$ is greater than both $\frac{q^n-1}{q^k-1}$ and $\frac{q^n-q^r}{q^k-1}$. So, in this case also, $\mbox{Orb}(U)$ cannot be a $0$-intersecting equidistant code. 

Hence, we conclude that a non-trivial $0$-intersecting equidistant single-orbit cyclic subspace code does not exist. 
\end{remark}
Next, we consider the $c$-intersecting $(c>0)$ equidistant single-orbit cyclic subspace codes.  

Consider an extension field $\mathbb{F}_{q^n}$ of $\mathbb{F}_q$. Let $\alpha$ be a primitive element of $\mathbb{F}_{q^n}$. Thus $\mathbb{F}_{q^n}^*=\{\alpha^{i}\mid i=0,1,\ldots,q^n-2\}$. Consider the group $(\mathbb{Z}_{q^n-1},\oplus_{q^n-1})$ of the integers modulo $q^n-1$ endowed with the addition modulo $q^n-1$. Define the map $ 	\Phi_{\alpha} :	(\mathbb{Z}_{q^n-1}, \oplus_{q^n-1})  \rightarrow (\mathbb{F}_{q^n}^*,\times)  $ by 
	\[i \rightarrow \alpha^i ~.\]

Then $\Phi_{\alpha}$  is a group isomorphism. In particular, for any subgroup 
  $G=\{\alpha^0=1, \alpha^{j_1},\alpha^{j_2},\ldots, \alpha^{j_m}\}$  of  $\mathbb{F}_{q^n}^*$, there exists a subgroup $I_G = \{t \mid \alpha^t \in G\}$ of $\mathbb{Z}_{q^n-1}$,
  and vice-versa, for a subgroup of $(\mathbb{Z}_{q^n-1}, \oplus_{q^n-1})$, there is a corresponding subgroup of $(\mathbb{F}_{q^n}^*,\times)$.

\begin{lemma}\label{lemma3}
Let $a, b, c$ and $m$ be positive integers such that $a < b \le c$. If $m^c-1$ divides $(m^a-1)(m^{b}-1)$, then  $c=b$.
\end{lemma}
\begin{proof}
	As $(m^c-1) \mid (m^a-1)(m^{b}-1)$, there exists a positive integer $k$ such that 
	
	\[(m^a-1)(m^{b}-1)=k (m^c-1)~.\]
	From this we get $m^{a+b}-m^a-m^{b}+1=k m^c- k$. This gives $m^a(m^{b}-1-m^{b-a}-k m^{c-a})=-(k+1)$, and hence $m^{a}$  divides $(k+1)$. Let $k+1=s m^a$ for some positive integer $s$. From this we get $k=s m^a-1$, and thus $(m^a-1)(m^{b}-1)= (s m^a-1) (m^c-1)$. Since $\frac{(m^b-1)}{(m^c-1)}\leq 1$, we get  $(sm^{a}-1)\leq (m^a-1)$. From this follows that $s=1$ and $c=b$. 
\end{proof}

\begin{lemma}\label{lemma4}
	Let $(G,+)$ be a group of order $v$ and $D\subseteq G$ with $\lvert D\rvert= k$. If for every $0\neq g\in G,~ \lvert D\cap (D+g)\rvert= \lambda>0$, then $D$ is a $(v,k,\lambda)$-difference set in $G$. 
\end{lemma}
\begin{proof}
	Suppose that the condition holds. Let $0\neq g\in G$ be an arbitrary element. As $\lvert D\cap (D+g)\rvert=\lambda$, there exist $\lambda$ number of distinct pairs $(d,d'),~ d,d'\in D$ such that 
	$d=d'+g$, i.e., $d-d'=g$. As $g$ is an arbitrary element of $G$, the multiset $[x-y: x, y \in D, x\neq y]$ contains every element in $G\backslash \{0\}$ exactly $\lambda$ times. Hence the result. 
\end{proof}
\begin{theorem}\label{thm2}
	Let $\alpha$ be a primitive element of $\mathbb{F}_{2^n}$ over $\mathbb{F}_2$. Let $U=\{0,\alpha^{i_1},\alpha^{i_2},$ $\ldots, \alpha^{i_{2^k-1}}\}$ be a subspace of dimension $k$ in $\mathbb{F}_{2^n}$ such that $U$ generates a full-length orbit.  Then the subspace code $\mbox{Orb}(U)$ is an  $r$-intersecting equidistant code $(r>0)$ if and only if the set of indices $i_j,~1\leq j\leq 2^k-1$, is a difference set in $\mathbb{Z}_{2^n-1}$. Moreover, the parameters of the difference set are $(2^n-1, 2^k-1, 2^r-1)$.
\end{theorem}
\begin{proof}
	Let $\mbox{Orb}(U)$ be an equidistant code and let $d_s(\mbox{Orb}(U))= 2(k-r)$, where $r>0$. Since $d_s(\mbox{Orb}(U))>0$, we have $0<r<k$. As $U$ generates a full-length orbit, for all $\beta\in \mathbb{F}_{2^n}\backslash \mathbb{F}_2$, $\dim(U\cap \beta U)=r$. Now consider the set $D=\{i_j\mid \alpha^{i_j}\in U\}$. Clearly $D\subseteq \mathbb{Z}_{2^n-1}$ and $\lvert D\rvert= 2^k-1$. Let $h(\neq0)$ be an arbitrary element in $\mathbb{Z}_{2^n-1}$. Then $\alpha^h\in \mathbb{F}_{2^n}\backslash \mathbb{F}_2$, and $\dim(U\cap \alpha^h U)=r$, i.e., $\lvert \{0, \alpha^{i_1},\alpha^{i_2},\ldots, \alpha^{i_{2^k-1}}\}\cap \{0, \alpha^{h+i_1 },\alpha^{h+i_2},\ldots, \alpha^{h+{i_{2^k-1}}}\}\rvert =2^r$. From this we get $\lvert D\cap (h+D)\rvert=2^r-1$. As $h$ is an arbitrary element in $\mathbb{Z}_{2^n-1}\backslash\{0\}$, by Lemma \ref{lemma4} we get the set of indices $D$ to be a $(2^n-1, 2^k-1, 2^r-1)$-difference set in $\mathbb{Z}_{2^n-1}$. 
	
	For the converse, let $D=\{i_j\mid \alpha^{i_j}\in U \}$ constitute a $(2^n-1, 2^k-1, s)$-difference set in $\mathbb{Z}_{2^n-1}$. From Equation (\ref{eqn1}), $s (2^n-2) = (2^k-1)(2^k-2)$. From this we get $s(2^{n-1}-1)= (2^k-1)(2^{k-1}-1)$. As $k<n$, from Lemma \ref{lemma3}, we get $k=n-1$, and so $s=(2^{k-1}-1)$. This implies that the multiset $[x-y: x,y\in D, x\neq y]$ contains every element of $\mathbb{Z}_{2^n-1}\backslash \{0\}$ exactly $2^{k-1}-1$ times.  Let $\alpha^m U\neq U$ be an arbitrary element in $\mbox{Orb}(U)$. Then $m\in \mathbb{Z}_{2^n-1}\backslash\{0\}$. By Lemma \ref{lemma1}, $\lvert D\cap (m+D)\rvert =2^{k-1}-1$. Therefore, $\lvert U\cap\alpha^m U\rvert= 2^{k-1}$ and $\dim(U\cap \alpha^m U)=k-1$. Hence $\mbox{Orb}(U)$ is an equidistant code. 
\end{proof}
\begin{remark}
	The argument used in Theorem \ref{thm2} cannot be applied for a subspace $U$ in $\mathbb{F}_{q^n}$ with $q>2$. Let $\alpha$ be a primitive element of $\mathbb{F}_{q^n}$ over $\mathbb{F}_q$  and let $U=\{0,\alpha^{i_1},\alpha^{i_2}, \ldots, \alpha^{i_{q^k-1}}\}$ be a subspace in $\mathbb{F}_{q^n}(q>2)$ over $\mathbb{F}_q$. Suppose that $U$ generates a full-length orbit. Then $\gamma U = U$ for any $\gamma\in \mathbb{F}_q^*$. Let  $D=\{i_j \mid \alpha^{i_j}\in U\}$. Note that for $2\in \mathbb{F}_q$, there exists an $h \in \mathbb{Z}_{q^n-1}$ such that $\alpha^h=2$. Now, on one hand, we have $\lvert D\cap(h+D)\rvert = q^r-1$, as $\mbox{Orb}(U)$ is $r$-intersecting code. On the other hand, for any $\alpha^{h'} \in \mathbb{F}_q$, $\lvert D\cap(h'+D)\rvert = q^k-1$. Thus, since $r < k,~ D$ is not a difference set in $G$.
\end{remark}

\begin{theorem}\label{thm3}
Let $\alpha$ be a primitive element of $\mathbb{F}_{q^n}$ over $\mathbb{F}_q$. Let $U=\{0,\alpha^{i_1},\alpha^{i_2},$ $\ldots,\alpha^{i_{q^k-1}}\}$ be a subspace in $\mathbb{F}_{q^n}$ of dimension $k$ such that $U$ generates a full-length orbit. If the subspace code $\mbox{Orb}(U)$ is a $r$-intersecting equidistant code $(r>0)$ then the indices $i_j$, $1\leq j\leq q^k-1$, form a $(q-1,\frac{q^n-1}{q-1}, q^k-1, q^k-1, q^r-1)$-relative difference set in $\mathbb{Z}_{q^n-1}$ . 
\end{theorem}
\begin{proof}
Let $\mbox{Orb}(U)$ be an equidistant subspace code, and let $d_s(\mbox{Orb}(U))=2(k-r)$, where $r>0$. Let $D=\{i_j\mid \alpha^{i_j}\in U\}$ and $N=\{h\mid \alpha^{h}\in \mathbb{F}_q^*\}$. Then $N$ is a subgroup of $\mathbb{Z}_{q^n-1}$ and $\lvert N\rvert= q-1$. For any $i\in \mathbb{Z}_{q^n-1}\backslash N,~ \alpha^i \in \mathbb{F}_{q^n}\backslash\mathbb{F}_q$ and thus $\dim(U\cap \alpha^iU)=r$. From this, we get $\lvert D\cap (i+D)\rvert=q^r-1$ for all $i \in \mathbb{Z}_{q^n-1}\backslash N$. Now for any $t\in N,~\alpha^t\in \mathbb{F}_q$ and $\dim(U\cap\alpha^t U)=q^k$. Thus, for any $t\in N, ~\lvert D\cap (t+D)\rvert=q^k-1$. Hence the set of indices $D$ constitutes a $(q-1,\frac{q^n-1}{q-1}, q^k-1, q^k-1, q^r-1)$ relative difference set in $\mathbb{Z}_{q^n-1}$ (relative to $N$). 
\end{proof}
\begin{remark}
If we take $q=2$ in Theorem \ref{thm3}, then subgroup $N=\{0\}$. In this case a relative differnce set is a difference set.  
\end{remark}

\begin{theorem}\label{thm4}
There is only the trivial equidistant full-length single-orbit cyclic subspace code  in $\mathcal{P}_q(n)$ for $ n\geq 3$.  
\end{theorem}
\begin{proof}
We have already proved in Remark \ref{newre} that if $U$ generates a full-length orbit, then there is no non-trivial $0$-intersecting single-orbit equidistant code.  
Let $\alpha$ be a primitive element of $\mathbb{F}_{q^n}$ over $\mathbb{F}_q$ and let $U=\{0,\alpha^{i_1}, \alpha^{i_2},$ $\ldots,\alpha^{i_{q^k-1}}\}$ be a subspace of dimension $k$ in $\mathbb{F}_{q^n}$ over $\mathbb{F}_q$. Let $\mbox{Orb}(U)$ be an equidistant subspace code with subspace distance $2(k-r)$, where $r>0$. Since $d_s(\mbox{Orb}(U))>0$, we have $0<r<k$. By Theorem \ref{thm3}, the set of indices $\{i_j\mid \alpha^{i_j}\in U\}$ constitutes a $(q-1,\frac{q^n-1}{q-1},q^k-1, q^k-1, q^r-1)$-relative difference set in $\mathbb{Z}_{q^n-1}$. By Lemma \ref{lemma2}, we get
\[(q^k-1)(q^k-2)=(q-1)\left(\frac{q^n-1}{q-1}-1\right)(q^r-1)+(q-2)(q^k-1)~.\] 
On simplifying the above equation, we get $(q^k-1)(q^{k-1}-1)= (q^{n-1}-1)(q^r-1)$. Further, this gives 
\begin{equation}\label{eqn2}
	q^{2k-1}-(q+1)q^{k-1}=q^{n+r-1}-q^{n-1}-q^r~. 
\end{equation}
 Suppose that $r<k-1$. On dividing both sides of Equation (\ref{eqn2}) by $q^{r}$, we get \[q^{2k-r-1}-(q+1)q^{k-r-1}=q^{n-1}-q^{n-r-1}-1~.\]
As $n>k>r+1$, the left side of the above equation is a multiple of $q$, but the right side is not. This is a contradiction.
So, we conclude that $r=k-1$. Substituting  $r=k-1$ in Equation (\ref{eqn2}), we get $k=n-1$. Therefore, $\dim(U)=n-1$ and $d_s(\mbox{Orb(U)})=2$. Hence the result.
\end{proof}
\begin{remark}
From the above theorem we conclude that for a subspace $U$ of $\mathbb{F}_{q^n}$ which generates a full-length orbit, $\mbox{Orb}(U)$ is an equidistant code if and only if $\dim(U)=1~\mbox{or}~n-1$. 
\end{remark}

Now, we consider the subspaces which do not generate a full-length orbit. The result of Theorem \ref{thm3} holds for such codes. This is shown in the following theorem.    

\begin{theorem}\label{thm5}
Let $\alpha$ be a primitive element of $\mathbb{F}_{q^n}$ over $\mathbb{F}_q$. Let $U=\{0,\alpha^{i_1},\alpha^{i_2},$ $\ldots,\alpha^{i_{q^k-1}}\}$ be a subspace in $\mathbb{F}_{q^n}$ of dimension $k$ such that $U$ does not generate a full-length orbit. If the subspace code $\mbox{Orb}(U)$ is a $r$-intersecting equidistant code $(r>0)$, then the indices $i_j$, $1\leq j\leq q^k-1$, form a  $(q^t-1, \frac{q^n-1}{q^t-1},q^k-1, q^k-1, q^r-1)$-relative difference set in $\mathbb{Z}_{q^n-1}$, where $t$ is a divisor of $\gcd(k,n)$.
\end{theorem}
\begin{proof}
Let $\mbox{Orb}(U)$ be an equidistant subspace code with subspace distance $2(k-r)$, where $r>0$. Let $\mbox{Stab}(U)=\mathbb{F}_{q^t}^*$ for some $t,~1<t<k$ and $t$ divides $\mbox{gcd}(k,n)$. Let $N=\{i_j\mid \alpha^{i_j}\in \mathbb{F}_{q^t}^*\}$. Then $N$ is a subgroup of $\mathbb{Z}_{q^n-1}$. Clearly, the cardinality of $N$ is $q^t-1$. Let $D=\{i_j\mid \alpha^{i_j}\in U\}$. For any $h \in N,~ U=\alpha^h U$. This gives $\lvert D \cap (h+D)\rvert =q^k-1$. For any $m\in \mathbb{Z}_{q^n-1}\backslash N,~ \dim(U\cap \alpha^m U)=r$. So, we get $\lvert D\cap(m+D)\rvert= q^r-1$. By Lemma \ref{lemma4}, the set of indices $D$ constitutes a $(q^t-1, \frac{q^n-1}{q^t-1},q^k-1, q^k-1, q^r-1)$-relative difference set in $\mathbb{Z}_{q^n-1}$.
\end{proof}  

\begin{theorem}\label{Theorem7}
There is only trivial equidistant single-orbit cyclic subspace code  in $\mathcal{P}_q(n)$ for $n\geq 3$.  
\end{theorem}
\begin{proof}
We have proved the result in Theorem \ref{thm4} for subspaces that generate the full-length orbit. Now we prove the result for subspaces that do not generate a full-length orbit. For $0$-intersecting equidistant subspace code, we have already proved the result in Remark \ref{newre}. Let $\alpha$ be a primitive element in $\mathbb{F}_{q^n}$, and let $U=\{0,\alpha^{i_1}, \alpha^{i_2},\ldots, \alpha^{q^k-1}\}$ be a subspace in $\mathbb{F}_{q^n}$ and let $\mbox{Stab}(U)=\mathbb{F}_{q^t}^*$. Let $D=\{i_j\mid \alpha^{i_j}\in U\}$. Let $\mbox{Orb}(U)$ be an equidistant subspace code with subspace distance $2(k-r) (r>0)$.  By Theorem \ref{thm5} the set of indices $D$ constitutes a $(q^t-1,\frac{q^n-1}{q^t-1},q^k-1,q^k-1,q^r-1)$-relative difference set in $\mathbb{Z}_{q^n-1}$. By Lemma \ref{lemma2}, 
\[(q^k-1)(q^k-2)=(q^t-1)\left(\frac{q^n-1}{q^t-1} -1\right)(q^r-1)+(q^t-2)(q^k-1)~.\]
As $k>t$, on simplifying the above equation, we get 
\begin{equation}\label{eqn3}
	q^{2k-t}-q^k-q^{k-t}=q^{n+r-t}-q^{n-t}-q^r~.
\end{equation}
\textbf{Case 1}. Let $r> k-t$. On dividing both sides of the Equation (\ref{eqn2}) by $q^{k-t}$, we get 
\[q^k-q^t-1=q^{n+r-k}-q^{n-k}-q^{r-k+t}~.\]
Clearly, the right side is a multiple of $q$ but the left side is not. This is a contradiction.\\
\textbf{Case 2}. Let $r< k-t$. On dividing both sides of Equation (\ref{eqn3}) by $q^r$, we get 
\[q^{2k-t-r}-q^{k-r}-q^{k-t-r}=q^{n-t}-q^{n-t-r}-1~.\]
As $n>k>t$ and $k>r$, the left side is a multiple of $q$ but the right side is not. This is a contradiction.

So, we conclude that $r=k-t$. Putting the  value of $r=k-t$ in Equation (\ref{eqn3}) we get $k=n-t$.  Thus, the dimension of subspace $U$ is $n-t$ and the subspace distance of $\mbox{Orb}(U)$ is $2t$. Hence the result. 
\end{proof}

\begin{remark}
From Theorem \ref{Theorem7}, it follows that a subspace $U$ in $\mathbb{F}_{q^n}$ with $\mbox{Stab}(U)=\mathbb{F}_{q^t}^*$, where $t\geq 1$, is equidistant if and only if the dimension of $U$ over $\mathbb{F}_q$ is $t~\mbox{or}~n-t$.
\end{remark}

\section{Equidistant single-orbit quasi-cyclic subspace codes}\label{sec4}

In this section we obtain some results on equidistant single-orbit quasi-cyclic subspace codes over $\mathbb{F}_q$.
\begin{definition}\cite[Definition 2.1]{glue}
	Fix an element $\beta\in \mathbb{F}_{q^n}^*\backslash\{1\}$. Let $U$ be an $\mathbb{F}_q$-subspace of $\mathbb{F}_{q^n}$. The $\beta$-cyclic orbit code generated by $U$ is defined as the set 
	\[\mbox{Orb}_\beta(U)=\{\beta^i U\mid i=0,1,\ldots,\lvert\beta\rvert-1\}~.\]
	If $\beta$ is a primitive element of $\mathbb{F}_{q^n}$, i.e., $\mathbb{F}_{q^n}^*=\langle\beta \rangle$, we write $\mbox{Orb}_\beta(U)$ simply as $\mbox{Orb}(U)$. 
\end{definition}
Let $G$ be a multiplicative subgroup of $\mathbb{F}_{q^n}^*$. A subspace code $C$ is called a \emph{quasi-cyclic} subspace code if $\alpha U\in C$ for all $U\in C$ and $\alpha \in G$ (see \cite{otal}). If $\beta\in \mathbb{F}_{q^n}^*\backslash\{1\}$ is not a primitive element of $\mathbb{F}_{q^n}$, we call $\mbox{Orb}_\beta(U)$ a single-orbit quasi-cyclic subspace code. The stabilizer of the subspace $U$ under the action of the group $\langle\beta\rangle$, denoted by $\mbox{Stab}_{\beta}(U)$, is defined as
\begin{eqnarray}\notag
	\mbox{Stab}_{\beta}(U)&=& \{ c \in \langle \beta\rangle \mid  c U =U\} \\ \label{new1}
	&=& \langle\beta\rangle \cap \mbox{Stab}(U)  ~, 
\end{eqnarray}
where $\mbox{Stab}(U)$ is the stabilizer of the subspace  $U$ under the action of the group $\mathbb{F}_{q^n}^*$. Clearly, $\mbox{Stab}_{\beta}(U)$ is a subgroup of $\langle \beta \rangle$. We can always find a positive divisor $s$ of $\lvert \beta \rvert$ such that $\mbox{Stab}_{\beta}(U)=\langle \beta^s\rangle$. From this, it follows that $\lvert \mbox{Stab}_{\beta}(U)\rvert =\frac{\lvert\beta\rvert}{s}$. Thus,
\begin{equation}\label{new2}
	\lvert\mbox{Orb}_{\beta}(U)\rvert =\frac{\lvert \beta\rvert}{\lvert \mbox{Stab}_{\beta}(U)\rvert}= s~.
	\end{equation}
From this, we see that the cardinality of the code $\mbox{Orb}_{\beta}(U)$ is a divisor of the order of $\beta$ and hence a divisor of the order of $\mathbb{F}_{q^n}^*$. 

In the previous section we have proved that there exist only trivial equidistant single-orbit cyclic subspace codes. Now, we turn to the $\beta$-cyclic orbit code $\mbox{Orb}_\beta(U)$, where $\beta$ is not a primitive element of $\mathbb{F}_{q^n}$. 
The following examples show that there exist equidistant single-orbit  quasi-cyclic subspace codes.

\begin{example}
	Consider the irreducible monic polynomial $p(x)=x^9+x^8+x^5+4x^4+x^3+2x^2+4x+3$ of degree $9$ over $\mathbb{F}_5$. Let $\alpha$ be a root of $p(x)$. Then $\mathbb{F}_5(\alpha)$ is an extension field of degree $9$ over $\mathbb{F}_5$. Consider the subspace $U=\delta_1 \mathbb{F}_5\oplus \delta_2 \mathbb{F}_5\oplus \delta_3\mathbb{F}_5\oplus\delta_4\mathbb{F}_5$, where $\delta_1=\alpha^6+2\alpha^2+1,~\delta_2=\alpha^5+\alpha^4+\alpha^2+2,~\delta_3=2\alpha^{4} + \alpha^3 + 2\alpha^2 + \alpha$ and $\delta_4=\alpha^{8} + \alpha^6 + \alpha^3$. The dimension of $U$ over $\mathbb{F}_5$ is $4$.  As $\gcd(\dim U,9)=1$, the stabilizer of  $U$ is $\mathbb{F}_5^*$. Thus, the subspace $U$ generates a full-length orbit. Let $\omega=2 \alpha^8+3\alpha^6+3\alpha^5+2\alpha^4+2\alpha^3$. The order of $\omega$ in $\mathbb{F}_{5^9}^*$ is $76$. Consider the orbit code $\mbox{Orb}_\omega(U)=\{\omega^i U \mid 1\leq i\leq \lvert \omega\rvert\}$. Using Magma Computational Algebra System (see \cite{magma}), we computed that $\dim(U\cap z U)=0$ for all $zU\in \mbox{Orb}_\omega(U),~ zU\neq U$, and $\lvert\mbox{Orb}_\omega(U)\rvert= 19$. Thus $\mbox{Orb}_\omega(U)$ is an equidistant code with subspace distance $8$.  
	
	Now let $G=\mathbb{F}_{5^3}$ and let $\beta=4 \alpha^{8} + \alpha^7 + \alpha^6 + \alpha^5 + 3\alpha^4 + 2\alpha^2 + 4\alpha + 3$, which is a generator of the group  $G\backslash\{0\}$. Consider the subspace code $\mbox{Orb}_\beta(U)=\{\beta^i U\mid 0 \leq i \leq \lvert \beta \rvert -1 \}$. Using  Magma, we computed that $\dim( U \cap c U)=0$ for all $c U \in \mbox{Orb}_\beta U,~ c U\neq U$, and $\lvert \mbox{Orb}_\beta(U)\rvert=31$. Hence $\mbox{Orb}_\beta(U)$ is an equidistant code with subspace distance $8$. 
\end{example}


\begin{example}\label{eg2}
	
	Consider the irreducible monic polynomial $p(x)= x^{12}+x^6+x^5+x^4+x^2+2$ of degree $12$ over $\mathbb{F}_3$. Let $\alpha$ be a root of $p(x)$. Then $\mathbb{F}_3(\alpha)$ is an extension field of degree $12$ over $\mathbb{F}_3$. Consider the subspace $U= \langle \alpha^{66430},\alpha^{199290},\alpha^{40880}, \alpha^{81760},\alpha^{286540}, \alpha^{374556}\rangle_{\mathbb{F}_3}$. The dimension of $U$ over $\mathbb{F}_3$ is $6$ and the subspace $U$ generates a full-length orbit. The subfield $\mathbb{F}_{3^3}$ of $\mathbb{F}_{3^{12}}$ is contained in $U$.  Let $c$ be a primitive element of $\mathbb{F}_{3^3}$. Then the order of $c$ is $26$. Consider the orbit code $\mbox{Orb}_c(U)=\{c^i U: i=1,\ldots, \lvert c \rvert\}$. Clearly $\mathbb{F}_{3^3}\subseteq U \cap c^i U$.  Using  Magma, we computed that $U \cap c^i U=\mathbb{F}_{3^3}$ for all $c^i U \in \mbox{Orb}_c(U)$ with $c^iU\neq U$. Thus $\mbox{Orb}_c(U)$ is an equidistant subspace code of size $13$ and minimum distance $6$.  Now, let $\gamma=\alpha^{9490}$. The order of $\gamma$ in $\mathbb{F}_{3^{12}}^*$ is $56$. Consider the orbit code $\mbox{Orb}_{\gamma}(U)=\{\gamma^i U: 1\leq i\leq \lvert \gamma \rvert\}$. By using Magma, we obtained that $\dim(U\cap \omega U)=2$ for all $\omega U \in \mbox{Orb}_{\gamma}(U)$ with $\omega U\neq U$. Thus $\mbox{Orb}_{\gamma}(U)$ is an equidistant code of size $28$ and minimum distance $8$. 
\end{example}
\begin{note}
	While considering an equidistant quasi-cyclic subspace code $\mbox{Orb}_{\beta}(U)$, we take $\beta\notin \mbox{Stab}(U)$. Otherwise, being a subfield of $\mathbb{F}_{q^n}$, the stabilizer of $U$ will contain the multiplicative subgroup of $\mathbb{F}_{q^n}^*$ generated by $\beta$, and  $\mbox{Orb}_{\beta}(U)$ will therefore contain a single subspace $U$. 
\end{note}

\begin{proposition}\label{prop2}
	Let $n$ be an even number and $U$ be a subspace in $\mathbb{F}_{q^n}$. Let $\beta$ be an element of degree $2$ in $\mathbb{F}_{q^n}$ such that $\beta \notin \mbox{Stab}(U)$. Then $\mbox{Orb}_\beta(U)$ is an equidistant subspace code. 
\end{proposition}
\begin{proof}
	
	Since $\beta \notin \mbox{Stab}(U)$, we have $\lvert \mbox{Orb}_\beta(U)\rvert \geq 2$. As $\beta$ is an element of degree $2$ in $\mathbb{F}_{q^n}$, $\mathbb{F}_q[\beta]=\{a+c\beta  \mid a,c \in \mathbb{F}_q\}$. Clearly, $\{\beta^i\mid 0\leq i\leq \lvert \beta \rvert -1\}\subseteq \mathbb{F}_q[\beta]$. By \cite[Lemma 2]{mah},  for any $\gamma\in \mathbb{F}_{q^n}^*$ and $s \in \mathbb{F}_q^*$, $\dim(U\cap \gamma U)= \dim(U\cap (\gamma+s)U)$.  
	Therefore, $\dim(U\cap\beta U)=\dim(U\cap\delta U)$ for all $\delta\in \mathbb{F}_q[\beta]\backslash\mathbb{F}_q$. Thus, $\dim(U\cap \beta U)=\dim(U\cap \beta ^i U)$ for all $i,~1\leq i\leq \lvert \beta \rvert -1$, with $\beta^iU \neq U$. Hence $\mbox{Orb}_\beta(U)$ is an equidistant code.  
\end{proof} 


\begin{definition}
	Let $U$ be a subspace of $\mathbb{F}_{q^n}$, and let $\beta\in \mathbb{F}_{q^n}^*\backslash \{1\}$.
	A $\beta$-cyclic orbit code $\mbox{Orb}_\beta(U)$ is called a sunflower if there exists a subspace $T$ of $\mathbb{F}_{q^n}$ such that for all $xU, zU \in \mbox{Orb}_\beta(U)$ with $xU\neq zU$ we have $xU\cap zU= T$. The subspace $T$, if it exists, is called the center of the sunflower $\mbox{Orb}_{\beta}(U)$. 
\end{definition}

If the center $T=\{0\}$, we say $\mbox{Orb}_{\beta}(U)$ is a sunflower with a trivial center. Clearly, a sunflower $\mbox{Orb}_\beta(U)$ is an equidistant code. Note that for an equidistant code $\mbox{Orb}_\beta(U)$, if there exists a subspace $S$ of $\mathbb{F}_{q^n}$ such that $U\cap xU=S$ for all $xU\in \mbox{Orb}_\beta(U)$, with $xU\neq U$, then $\mbox{Orb}_\beta(U)$ is a sunflower.

In what follows, we give an example of a sunflower in a field extension of degree $15$ over $\mathbb{F}_3$.  
\begin{example}
	Consider the irreducible monic polynomial $p(x)=x^{15} + 2x^8 + x^5 + 2x^2 + x + 1$ of degree $15$ over $\mathbb{F}_3$. Let $\alpha$ be a root of $p(x)$. Then $\mathbb{F}_3(\alpha)\simeq \mathbb{F}_{3^{15}}$. Let $\rho =\alpha^{13} + 2\alpha^{12} + \alpha^9 + \alpha^8 + 2\alpha^6 + 2\alpha^5 + \alpha^3 + 2\alpha + 1$. The degree of the minimal polynomial of $\rho$ over $\mathbb{F}_3$ is 3. Then $\mathbb{F}_{3^3}=\langle 1,\rho, \rho^2\rangle_{\mathbb{F}_3}$.  Let $U=\beta_1\mathbb{F}_3^3\oplus \beta_2\mathbb{F}_3\oplus \beta_3\mathbb{F}_3$, where $\beta_1=\alpha^7,~\beta_2=2\alpha^{14} + 2\alpha^{12} + \alpha^9 + 2\alpha^6 + 2\alpha^5 + \alpha^3 + \alpha^2 + 2\alpha$ and $\beta_3=\alpha^{14} + \alpha^{13} + \alpha^{11} + \alpha^9 + 2\alpha^8 + 2\alpha^7 + 2\alpha^6 + \alpha^5 + \alpha^4 + 2$. The dimension of $U$ over $\mathbb{F}_3$ is $5$, and $U$ generates a full-length orbit. Let $G_1=\mathbb{F}_{3^3}$. The order of $\rho$ is $26$. Therefore $\rho$ is a generator of the multiplicative group $G_1\backslash\{0\}$. Consider the subspace code $\mbox{Orb}_\rho(U)=\{\rho^i U\mid i=0,1,\ldots, \lvert \rho \rvert-1\}$. By using Magma, we obtained that $\dim(U\cap x U)=3$ and $U \cap x U =\delta_1\mathbb{F}_3\oplus \delta_2 \mathbb{F}_3 \oplus \delta_3 \mathbb{F}_3$, where $\delta_1= \alpha^{12}+2\alpha^9+\alpha^7+2\alpha^5+\alpha^4+\alpha^3+\alpha+2,~ \delta_2=\alpha^{12}+2\alpha^9+2\alpha^5+\alpha^4+\alpha^3+\alpha+2,~ \delta_3=\alpha^{14}+\alpha^{13}+2\alpha^{12}+2\alpha^{11}+\alpha^8+2 \alpha^7+2\alpha^6+2\alpha^4+2\alpha^2$ for all $xU\in \mbox{Orb}_\rho(U)$ with $xU\neq U$. Thus, $\mbox{Orb}_\rho(U)$ is a sunflower of size $13$ and the subspace distance $4$.

\end{example} 
\begin{note}
	 In  Example \ref{eg2}, we have $U\cap \gamma U=\langle \alpha^{32120}, \alpha^{91250} \rangle_{\mathbb{F}_3}$ and  $U \cap \gamma^2 U=\langle \alpha^{91250}, \alpha^{143080}  \rangle_{\mathbb{F}_3}$. Using Magma, we found that $U\cap \gamma U$ and  $U \cap \gamma^2 U$ are not equal. Thus, $\mbox{Orb}_{\gamma}(U)$ is an equidistant code, but not a sunflower.
\end{note}	

In Proposition \ref{prop2}, we have proved that if $\beta$ is an element of degree $2$ in  $\mathbb{F}_{q^n}$,  then a $\beta$-cyclic orbit code is equidistant. Next we prove that it is indeed a sunflower.  
\begin{theorem}\label{thm6}
	Let $n$ be an even number and $U$ be a subspace of $\mathbb{F}_{q^n}$. For any element $\beta$ of degree $2$ in $\mathbb{F}_{q^n}$ with $\beta \notin \mbox{Stab}(U)$, $\mbox{Orb}_\beta(U)$ is a sunflower. 
\end{theorem}
\begin{proof}
	Let $\beta$ be an element of degree $2$ in $\mathbb{F}_{q^n}$ such that $\beta \notin \mbox{Stab}(U)$. \\
	\textbf{Case 1}. Let $\dim(U\cap \beta U)=0$. By Proposition \ref{prop2}, $\mbox{Orb}_\beta(U)$ is an equidistant code.  Then $U\cap \beta^i U=\{0\}$ for all $i\in \{0,1,\ldots, \lvert \beta\rvert-1\}$ with $\beta^iU\neq U$. Thus $\mbox{Orb}_\beta(U)$ is a sunflower. \\
	\textbf{Case 2}. Let $\dim(U\cap \beta U)\neq 0$ and let $V=U\cap \beta U$. By \cite[Theorem 5]{mah}, $\mathbb{F}_{q^2}^*\subseteq \mbox{Stab}(V)$. Now, consider an element $\beta^j,~ j\in \{0,1,\ldots, \lvert \beta \rvert -1\}$, such that $\beta^jU\neq U$. Then $ \beta^j \in \mathbb{F}_q[\beta]\backslash\mathbb{F}_q$. Let $\beta^j=a\beta +c$ for some $a,c\in \mathbb{F}_q$ and $a \neq 0$. As $\mathbb{F}_{q^2}^*\subseteq \mbox{Stab}(V),~ (a\beta+c)^{-1}V=V$. Thus $(a\beta+c)^{-1}V\subseteq U$ and  $V\subseteq U\cap (a\beta+c)U$. As $\mbox{Orb}_\beta(U)$ is an equidistant code, $\dim(U\cap\beta U)=\dim(U\cap (a\beta+c)U)$. So, we get $V=  U\cap (a\beta+c)U = U\cap \beta^jU$. As $\beta^j$ is an arbitrary element, $V=U\cap \beta^i U$ for all $i\in \{0,1,\ldots,\lvert\beta\rvert-1\}$ with $\beta^iU\neq U$. Hence, $\mbox{Orb}_{\beta}(U)$ is a sunflower. 
\end{proof}


\begin{remark}
	Let $n$ be an even number, and let $G=\mathbb{F}_{q^2}^*$, so that G is a multiplicative subgroup of $\mathbb{F}_{q^n}^*$. Let $\gamma$ be a generator of $G$, i.e., $G=\{\gamma^i\mid 0\leq i \leq q^2-2\}$. Clearly, degree of $\gamma$ over $\mathbb{F}_q$ is $2$. If $\mathbb{F}_{q^2}^*\nsubseteq \mbox{Stab}(U)$, by Theorem \ref{thm6}, $\mbox{Orb}_\gamma(U)$ is a sunflower. 
\end{remark}

\begin{proposition}
	Let $U$ be a subspace of dimension $k~\mbox{in}~\mathbb{F}_{q^n}$ such that $U$ generates a full-length orbit. Let $V$ be a subspace of $U$ of dimension $k-1$ over $\mathbb{F}_q$ and $\mbox{Stab}(V)=\mathbb{F}_{q^t}^*~(t>1)$.  Then the orbit code  $\{\alpha U\mid \alpha \in \mathbb{F}_{q^t}^*\}$ is a sunflower.
\end{proposition}
\begin{proof}
	We can write $U=V\oplus \langle x\rangle_{\mathbb{F}_q}$ for some $x\in \mathbb{F}_{q^n}^*$. Let $\omega\in \mathbb{F}_{q^t}^*\backslash \mathbb{F}_q$. As $\omega V=V,~\omega U= V\oplus \langle \omega x\rangle_{\mathbb{F}_q}$. From this we get  $V\subseteq U\cap \omega U$. Thus, $\dim(U\cap \omega U)\geq k-1$. If $\dim(U\cap \omega U)=k$ then $\omega \in \mbox{Stab}(U)$. This is a contradiction as $\mbox{Stab}(U)=\mathbb{F}_q^*$. So, we get $U\cap \omega U=V$.  Since $\omega$ is an arbitrary element in $\mathbb{F}_{q^t}^*\backslash\mathbb{F}_q$, we have $U\cap \beta  U=V$ for all $\beta \in \mathbb{F}_{q^t}\backslash\mathbb{F}_q$. Hence the result. 
\end{proof}

\begin{proposition}\label{prop4}
	For any sunflower $\mbox{Orb}_{\beta}(U)$, with $\beta \notin \mbox{Stab}(U)$, the center does not generate a full-length orbit. 
\end{proposition}
\begin{proof}
	Let $U$ be a subspace of $\mathbb{F}_{q^n}$, and let $\beta\in \mathbb{F}_{q^n}\backslash \mbox{Stab}(U)$ be such that $\mbox{Orb}_{\beta}(U)$ is a sunflower. Let $V$ be the center of the sunflower $\mbox{Orb}_{\beta}(U)$. If $V=\{0\}$ then the result is trivially true. Let $V\neq\{0\}$. As $V=U\cap \beta U$,  we have $\beta V=\beta U\cap\beta^2 U$. If $\beta ^2\in \mathbb{F}_q$, then $\beta^2U=U$, and so $V=\beta V$. Now, let $\beta^2\notin \mathbb{F}_q$. Since $V$ is the center of $\mbox{Orb}_{\beta}(U)$, $V=U\cap \beta U=U\cap \beta^2U$. Thus, $V\subseteq \beta U \cap \beta^2 U=\beta (U\cap \beta U)=\beta V$. From this we get $V=\beta V$. Hence, $\beta \in \mbox{Stab}(V)$. The result follows.
\end{proof}

\begin{remark}\label{remark5}
	By Proposition \ref{prop4}, for a sunflower $\mbox{Orb}_{\beta}(U)~(\beta \notin \mbox{Stab}(U))$  with  center $V\neq\{0\},~ \beta\in \mbox{Stab}(V)$. It is known that $\mbox{Stab}(V)$ is a subgroup of $(\mathbb{F}_{q^n}^*,\times)$. So, we conclude that $\{\beta^i\mid i=0,1,\ldots,\lvert \beta\rvert-1\}\subseteq \mbox{Stab}(V)$. 
\end{remark}
\begin{remark}
We can quickly check that a subspace of dimension one generates a full-length orbit. Thus, according to Proposition \ref{prop4}, the dimension of the non-trivial center of a sunflower quasi-cyclic orbit code is always greater than or equal to two. However, $1$-intersecting equidistant quasi-cyclic orbit codes which are not sunflower can exist in $\mathbb{F}_{q^n}$. Next, we provide an example of such a code.
\end{remark}

\begin{example}
Consider the monic irreducible polynomial $p(x)= x^{10} + x^6 + x^5 + x^3 + x^2 + x + 1$ of degree $10$ over $\mathbb{F}_2$. Let $\alpha$ be a root of $p(x)$. Then $\mathbb{F}_2(\alpha)$ is an extension field of degree $10$ over $\mathbb{F}_2$. Let $U=\langle 1, \alpha^{13}, \alpha^{70}, \alpha^{177}\rangle_{\mathbb{F}_2} $. The dimension of $U$ over $\mathbb{F}_2$ is $4$.   The cardinality of the code $\mbox{Orb}(U)=\{\gamma U \mid \gamma\in \mathbb{F}_{2^{10}}^*\}$ is $1023$. From this follows that $U$ generates a full-length orbit. Let $\beta=\alpha^{93}$ be an element of order $11$ in $\mathbb{F}_{2^{10}}^*$. By using Magma, we get that $\dim(U\cap \beta ^i U)= 1$ for all $i$ in $\{0,1,\ldots, \rvert\beta\lvert-1\}$ with $\beta^iU\neq U$. Thus, $\mbox{Orb}_{\beta}(U)$ is a $1$- intersecting equidistant code. As $U \cap \beta U=\{0, \alpha^{457}\}$ and $U\cap \beta^2 U=\{0, \alpha^{415}\}$, $\mbox{Orb}_{\beta}(U)$ is not a sunflower. 
	
\end{example}

\begin{corollary}
	Let $U$ be a subspace of dimension $k$ in $\mathbb{F}_{q^n}$ such that $U$ generates a full-length orbit. Let $\beta$ be an element in  $\mathbb{F}_{q^n}\backslash \mathbb{F}_q$ such that $\mbox{Orb}_{\beta}(U)$ is a sunflower. Let $V$ be the center of the sunflower and let $\mbox{Stab}(V)=\mathbb{F}_{q^t}^*$. If the dimension of $V$ is $k-1$, then $\mathbb{F}_{q^t}^*$  is the largest subgroup of $\mathbb{F}_{q^n}^*$ such that $\{\gamma U\mid \gamma \in \mathbb{F}_{q^t}^*\}$ is a sunflower with the center $V$.  
\end{corollary}
\begin{proof}
	Let the dimension of $V$ be $k-1$, and let $x$ be any arbitrary element in $\mathbb{F}_{q^t}\backslash \mathbb{F}_q$. As $\mathbb{F}_{q^t}^*$ is the stabilizer of $V$,  $x V =V$.  Thus, $V\subseteq U \cap xV\subseteq  U\cap x U$. The dimension of $V$ is $k-1$ and $U$ generates a full-length orbit. So, we get $V=U\cap x U$. Since $x$ is an arbitrary element in $\mathbb{F}_{q^t}\backslash\mathbb{F}_q$, we get $V=U\cap x U$ for all $x\in \mathbb{F}_{q^t}^*$ with $x U\neq U$. Let $\delta$ be a generator of the multiplicative group $\mathbb{F}_{q^t}^*$. Then, $\mbox{Orb}_{\delta}(U)=\{\delta^i U\mid i=0,1,\ldots, \lvert\delta\rvert-1 \}$ is a sunflower with the center $V$. By Remark \ref{remark5}, for any $\beta\in \mathbb{F}_{q^n}^*$ such that $\mbox{Orb}_{\beta}(U)$ is a sunflower with center $V$,   $\{\beta^i \mid i=0,1,\ldots, \lvert\beta\rvert-1\}\subseteq \mathbb{F}_{q^t}^*$.  Hence the result. 
	\end{proof}
\begin{corollary}
	If $n$ is a prime number, then there does not exist a sunflower $\mbox{Orb}_{\beta}(U)$ in $\mathbb{F}_{q^n}$ with a non-trivial center.
\end{corollary}
\begin{proof}
By \cite[Theorem 1]{otal}, a subspace $W$ of $\mathbb{F}_{q^n}$ does not generate a full-length orbit if and only if its stabilizer is $\mathbb{F}_{q^t}^*$ for some $t>1$ dividing $\mbox{gcd}(\dim(W),n)$. Thus, for a prime number $n$, every proper subspace in $\mathbb{F}_{q^n}$ generates a full-length orbit. Now, it is clear by Proposition \ref{prop4} that for a prime number $n$, there does not exist any sunflower  $\mbox{Orb}_{\beta}(U)$  with a non-trivial center in $\mathbb{F}_{q^n}$. 
\end{proof}
A sunflower with a trivial center may exist in $\mathbb{F}_{q^n}$ for a prime $n$. We give below such an example. 
\begin{example}
	Consider the monic irreducible polynomial $p(x)=x^{11}+2x^2+1$ of degree $11$ over $\mathbb{F}_3$. Let $\alpha$ be a root of $p(x)$. Then, $\mathbb{F}_3(\alpha)$ is an extension field of degree $11$ over $\mathbb{F}_3$. Let $U=\langle \alpha^{3179}, \alpha^{8390}, \alpha^{31874},\alpha^{114951}, \alpha^{118325} \rangle_{\mathbb{F}_3}$.  The dimension of $U$ over $\mathbb{F}_3$ is $5$, and $U$ generates a full-length orbit. Let $c=z^{3851}$ be an element in $\mathbb{F}_3(\alpha)$. The multiplicative order of $c$ is $46$. Using Magma, we obtained that $U\cap c^i U=\{0\}$ for all $i$ in $\{0,1,\ldots, \lvert c\rvert-1\}$ with $c^iU\neq U$. Thus, $\mbox{Orb}_c(U)$ is a sunflower with a trivial center.  The cardinality of $\mbox{Orb}_c(U)$ is  $23$, and the subspace distance is $10$. 	
\end{example}

The following theorem gives an upper bound on the cardinality of a sunflower in $\mathbb{F}_{q^n}$ with a non-trivial center.

\begin{theorem}\label{thm7}
	Let $U$ be a subspace of dimension $k$ in $\mathbb{F}_{q^n}$ such that $U$ generates a full-length orbit. Let $\mbox{Orb}_{\beta}(U)~(\beta\in \mathbb{F}_{q^n}\backslash\mathbb{F}_q)$ be a sunflower with a non-trivial center. Then 
	\[\lvert \mbox{Orb}_{\beta}(U)\rvert \leq \frac{q^s-1}{q-1}~,\]
	where $s$ is the largest positive divisor of $n$ such that $s<k$.
\end{theorem}
\begin{proof}
	Let $V$ be the center of the sunflower $\mbox{Orb}_{\beta}(U)~(\beta\in \mathbb{F}_{q^n}\backslash\mathbb{F}_q)$  such that $V\neq \{0\}$. Then, $V=U\cap\beta^iU$ for all $i\in \{0,1,\ldots,\lvert\beta\rvert-1\}$ with $\beta^i U\neq U$.  Since the dimension of $U$ is $k$, the dimension of $V$ is less than or equal to $k-1$. By Remark \ref{remark5}, $\{\beta^i \mid i=0,1,\ldots,\lvert\beta\rvert-1\}\subseteq\mbox{Stab}(V)$. As $\mbox{Stab}(V)\cup\{0\}$ is a subfield of $\mathbb{F}_{q^n}$ and $V$ is a vector space over $\mbox{Stab}(V)\cup\{0\}$, we have  $\mbox{Stab}(V)=\mathbb{F}_{q^s}^*$ for some positive integer $s>1$ dividing $\mbox{gcd}(\dim(V),n)$. From this follows that $s\leq k-1$. Let $\mbox{Stab}(V)=\mathbb{F}_{q^s}^*=\langle \delta \rangle$ for some $\delta\in \mathbb{F}_{q^n}$. Since $\{\beta^i\mid i=0,1,\ldots,\lvert\beta\rvert-1\}\subseteq \mathbb{F}_{q^s}^*$, $\mbox{Orb}_{\beta}(U) \subseteq \mbox{Orb}_{\delta}(U)$. As $U$ generates a full-length orbit, by Equations (\ref{new1}) and (\ref{new2}), we get $\lvert\mbox{Orb}_{\beta}(U)\rvert\leq \lvert\mbox{Orb}_{\delta}(U)\rvert= \frac{q^s-1}{q-1}$. Hence the result. 
\end{proof}

Let $U$ be a subspace in $\mathbb{F}_{q^n}$ of dimension $k$ such that $U$ does not generate a full-length orbit, and let $\mbox{Stab}(U)=\mathbb{F}_{q^t}^*$. Let $\gamma\in \mathbb{F}_{q^n}\backslash\mathbb{F}_{q^t}$ and let $\mbox{Orb}_{\gamma}(U)$ be a sunflower with a non-trivial center $V$. 
Then $V=U\cap \gamma U$. For any $a \in \mathbb{F}_{q^t}^*,~ a V= a U \cap a\gamma  U$. As $a\in \mbox{Stab}(U)$, $a U=U$ and $a \gamma U=\gamma U$. Thus, $a V=V$.  From this follows that $a\in \mbox{Stab}(V)$. Since $a$ is an arbitrary element in $\mbox{Stab}(U)$, we get  $\mbox{Stab}(U)\subseteq \mbox{Stab}(V)$. Let $\mbox{Stab}(V)=\mathbb{F}_{q^s}^*=\langle \delta \rangle$ for some positive integer $s>1$ dividing $\mbox{gcd}(\dim(V),n)$ and $\delta \in \mathbb{F}_{q^n}$. Note that by Remark \ref{remark5}, $\langle \gamma \rangle \subseteq \langle \delta \rangle$, thus $\mbox{Orb}_{\gamma}(U) \subseteq \mbox{Orb}_{\delta}(U)$. So, by Equations (\ref{new1}) and (\ref{new2}), we get  
\[\lvert \mbox{Orb}_{\gamma}(U)\rvert \leq \lvert \mbox{Orb}_{\delta}(U)\rvert=\frac{\lvert \delta \rvert} {\lvert \langle \delta \rangle \cap \mathbb{F}_{q^t}^* \rvert}=\frac{q^s-1}{q^t-1}~.\] 
Hence 
\[\lvert\mbox{Orb}_{\gamma}(U)\rvert\leq \frac{q^s-1}{q^t-1}~,\]
where $s$ is the largest positive divisor of $n$ such that $s<k$ and $t \mid s$.

The sunflower constructed in the following example is optimal. 
\begin{example}
	Consider the monic irreducible polynomial $p(x)=x^{15} + 2x^5 + 3x^3 + 3x^2 + 4x + 3$ of degree $15$ over $\mathbb{F}_5$. Let $\alpha$ be a root of $p(x)$. Then $\mathbb{F}_5(\alpha)$ is an extension field of degree $15$ over $\mathbb{F}_5$ and $\mathbb{F}_5(\alpha)\simeq\mathbb{F}_{5^{15}}$. Let $\beta= 3\alpha^{14} + \alpha^{12} + 4\alpha^{11} + 4\alpha^{10} + 2\alpha^9 + \alpha^8 + 4\alpha^7 + \alpha^6 + 4\alpha^5 + 2\alpha^4+ 2\alpha + 2$ be an element in $\mathbb{F}_{5^{15}}^*$. The order of $\beta$ in $\mathbb{F}_{5^{15}}^*$ is $5^5-1=3124$. Then, $\beta$ is a generator of the multiplicative subgroup $\mathbb{F}_{5^5}^*$ of $\mathbb{F}_{5^{15}}^*$. Let $U=(\alpha^2+1)\mathbb{F}_{5^5}\oplus (\alpha^7+2\alpha+3)\mathbb{F}_5\oplus (\alpha^9+2\alpha^2+3)\mathbb{F}_5$. The dimension of $U$ over $\mathbb{F}_5$ is $7$, and the subspace $U$ generates a full-length orbit. Now consider the subgroup $\mbox{Orb}_{\beta}(U)=\{\beta^i U \mid 0\leq i \leq \lvert\beta\rvert-1\}$. By using Magma, we computed that $U\cap \beta^i U=\mathbb{F}_{5^5}$ for all $i$ in $\{1,2,\ldots,\lvert\beta\rvert-1\}$ with $\beta^iU\neq U$. Thus, $\mbox{Orb}_{\beta}(U)$ is a sunflower with the center $\mathbb{F}_{5^5}$. Using Magma, we computed that $\lvert\mbox{Orb}_{\beta}(U)\rvert = \frac{5^5-1}{5-1}=781$. As $5~(<k=7)$ is the largest positive divisor of $15$, the cardinality of $\mbox{Orb}_{\beta}(U)$ is maximum.  
\end{example}

The cardinality of a sunflower $\mbox{Orb}_{\beta}(U)$ in $\mathbb{F}_{q^n}$ with a trivial center may be greater than $\frac{q^s-1}{q-1}$, where $s$ is the largest positive divisor of $n$ such that $s<\dim(U)$. The following example illustrates this. 
\begin{example}
	Consider the monic irreducible polynomial $p(x)=x^{12} + x^6 + x^5 + x^4 + x^2 + 2$ of degree $12$ over $\mathbb{F}_3$. Let $\alpha$ be a root of $p(x)$. Then, $\mathbb{F}_3(\alpha)$ is an extension field of degree $12$ over $\mathbb{F}_3$. Let $U=\langle\alpha^{565},\alpha^{123982},\alpha^{179292}, \alpha^{208314},\alpha^{395390} \rangle_{\mathbb{F}_3}$. The dimension of $U$ over $\mathbb{F}_3$ is $5$, and $U$ generates a full-length orbit. Let $\gamma=\alpha^{4088}$ be an element in $\mathbb{F}_{3^{12}}$. The multiplicative order of $\gamma$ is $130$. By using Magma, we computed that $U\cap\gamma^iU=\{0\}$ for all $i$ in $\{1,\ldots,\lvert\gamma\rvert\}$ with $\gamma^iU\neq U$. Thus, $\mbox{Orb}_{\gamma}(U)$ is a sunflower with a trivial center. The cardinality of $\mbox{Orb}_\gamma(U)$ is $65$. Here we have $n=12$ and $k=5$. So the largest divisor of $n$ less than $k$ is $4$. Clearly, $\lvert\mbox{Orb}_{\gamma}(U)\rvert =65 > \frac{3^4-1}{3-1}=40$. 
\end{example}

Now we  discuss about the maximum size of a sunflower with a trivial center. 

Let $U$ be a subspace of dimension $k~ \mbox{in}~ \mathbb{F}_{q^n}$ such that $\mbox{Orb}_{\beta}(U)$ is a sunflower with a trivial center.

\noindent \textbf{Case 1}. If $k$ divides $n$ then
$\mbox{Orb}_{\beta}(U)$ is clearly a subset of a $k$-spread. Therefore, by Theorem \ref{thm1},
\[\lvert\mbox{Orb}_{\beta}(U)\rvert\leq \frac{q^n-1}{q^k-1}~.\]
This bound is achievable. The following example illustrates this. 
\begin{example}
	Consider the monic irreducible polynomial $p(x)=x^{12} + x^7 + x^6 + x^5 + x^3 + x + 1$ of degree $12$ over $\mathbb{F}_2$. Let $\alpha$ be a root of $p(x)$. Then $\mathbb{F}_2(\alpha)$ is an extension field of degree $12$ over $\mathbb{F}_2$, and $\mathbb{F}_2(\alpha)\simeq \mathbb{F}_{2^{12}}$. Let $U=\langle 1, \alpha^{470}, \alpha^{3607},\alpha^{3621} \rangle_{\mathbb{F}_2} $. The dimension of $U$ over $\mathbb{F}_2$ is $4$, and $U$ generates a full-length orbit. Let $\gamma=\alpha^{15}$. The multiplicative order of $\gamma$ in $\mathbb{F}_{2^{12}}^*$ is $273$. Consider the orbit code $\mbox{Orb}_{\gamma}(U)=\{\gamma^i U \mid 0\leq i \leq \lvert \gamma \rvert-1\}$. Using Magma, we computed that $U\cap x U=\{0\}$ for all $xU\in \mbox{Orb}_{\gamma}U$, with $xU\neq U$.  Thus, $\mbox{Orb}_{\gamma}(U)$ is a sunflower with a trivial center. 
	 The computation through Magma shows that the cardinality of $\mbox{Orb}_{\gamma}(U)$ is $273$, which is equal to $\frac{2^{12}-1}{2^4-1}$. Hence $\mbox{Orb}_{\gamma}(U)$ is an optimal sunflower with a trivial center. 
	\end{example}

\noindent \textbf{Case 2}. If $k$ does not divide $n$ then $\mbox{Orb}_{\beta}(U)$ is a subset of a partial $k$-spread. Let $r$ denote the remainder when $n$ is divided by $k$. Then, by Lemma \ref{lem5}, we get 
\[\lvert\mbox{Orb}_{\beta}(U)\rvert \leq \frac{q^n-q^r}{q^k-1}~.\]
From this, it follows that $\lvert \mbox{Orb}_{\beta}(U)\rvert \leq\frac{q^r(q^{n-r}-1)}{q^k-1}$. We know that the cardinality of $\mbox{Orb}_{\beta}(U)$ is a divisor of the order of $\mathbb{F}_{q^n}^*$. However, $ \frac{q^r(q^{n-r}-1)}{q^k-1}$ does not divide $q^n-1$.  Hence, in this case, $\lvert\mbox{Orb}_{\beta}(U)\rvert< \frac{q^n-q^r}{q^k-1}$.

\section{Conclusion}
 In this paper, we explored equidistant single-orbit cyclic and quasi-cyclic subspace codes. We proved that there are no non-trivial equidistant single-orbit cyclic subspace codes. Furthermore, we presented examples of equidistant single-orbit quasi-cyclic subspace codes and discussed the largest cardinality of a sunflower. However, more investigation is needed regarding the cardinality of a single-orbit quasi-cyclic subspace code. For a given extension field  $\mathbb{F}_{q^n}$, the largest cardinality of a $c$-intersecting equidistant orbit code in $\mathbb{F}_{q^n}$ remains unknown. Future research could focus on constructing the largest $c$-intersecting equidistant orbit codes and sunflowers. 

\bmhead{Data Availability}
No data was used for the research described in this article.

\bmhead{Conflicts of interest}
The authors have no conflicts of interest to declare that are relevant to the content of this article.

\bmhead{Acknowledgments}
The authors would like to thank the anonymous referees for their careful reading of the manuscript and valuable comments, which greatly improved the final presentation of the paper. Also, the first author would like to thank Ministry of Education, India for providing financial support.  
\bibliography{main}
\end{document}